\documentclass[prb,twocolumn,superscriptaddress,showpacs]{revtex4}
\usepackage{graphicx}
\usepackage{color}
\usepackage{amssymb,amsmath}

\begin{document}

\title{Unidirectional $d$-wave superconducting domains in the 
two-dimensional t-J model}

\author{     Marcin Raczkowski}
\affiliation{Marian Smoluchowski Institute of Physics, Jagellonian
             University, Reymonta 4, PL-30059 Krak\'ow, Poland}

\author{     Manuela Capello}
\affiliation{Laboratoire de Physique Th\'eorique UMR 5152, CNRS and
             Universit\'e Paul Sabatier, F-31062 Toulouse, France}

\author{     Didier Poilblanc}
\affiliation{Laboratoire de Physique Th\'eorique UMR 5152, CNRS and
             Universit\'e Paul Sabatier, F-31062 Toulouse, France}

\author{     Raymond Fr\'esard}
\affiliation{\mbox{Laboratoire CRISMAT, UMR 6508 CNRS--ENSICAEN, 
             6 Bld. du Mar\'echal Juin, F-14050 Caen, France}}

\author{     Andrzej M. Ole\'s}
\affiliation{Marian Smoluchowski Institute of Physics, Jagellonian
             University, Reymonta 4, PL-30059 Krak\'ow, Poland}
\affiliation{Max-Planck-Institut f\"ur Festk\"orperforschung,
             Heisenbergstrasse 1, D-70569 Stuttgart, Germany }

\date{\today}

\begin{abstract}
Motivated by the recently observed pattern of unidirectional domains in 
high-$T_c$ superconductors [Y. Kohsaka {\it et al.}, Science {\bf 315}, 
1380 (2007)], we investigate the emergence of spontaneous modulations 
in the $d$-wave superconducting resonating valence bond phase using 
the $t$-$J$ model at $x=1/8$ doping.
Half-filled charge domains separated by four lattice 
spacings are found to form along one of the crystal axis leading to modulated
superconductivity with {\it out-of-phase\/} $d$-wave order parameters in
neighboring domains. Both renormalized mean-field theory and variational 
Monte Carlo calculations yield that the energies of modulated and uniform 
phases are very close to each other.
\end{abstract}

\maketitle

Puzzling properties of the high-$T_c$ superconductors have often
been attributed to competing instabilities.\cite{Voj00} Indeed,
it is believed that doping an antiferromagnetic (AF) Mott insulator
(which could be described e.g. by the so-called $t$-$J$ model\cite{tJ}) 
leads to quantum disordered states with short-ranged magnetic
correlations between $S=1/2$ spins and exotic properties.\cite{Suc03} 
Among them, the resonating valence bond (RVB) state was the first 
theoretical proposal supposed to capture the essence of high-$T_c$ 
superconductivity.\cite{And87}
Remarkably, this approach based on the Gutzwiller-projected BCS
trial wavefunction, the parameters of which are usually determined either
by using renormalized mean-field theory\cite{RVB} (RMFT) or by
variational Monte Carlo (VMC) method,\cite{VMC} not only correctly
predicted  the $d$-wave symmetry of the superconducting 
(SC) order parameter,\cite{dwave} but reproduced in addition the 
experimentally observed doping dependence of a variety of physical
observables in the SC regime.\cite{th_exp}

In fact, exactly at half-filling where a particle-hole SU(2) symmetry applies, 
the RVB phase is equivalent to the staggered flux (SF) state,\cite{sfp} 
a projected Slater determinant build from a tight binding model under 
a staggered magnetic flux. Remarkably, short-ranged staggered orbital current 
correlations have been seen in the Gutzwiller-projected $d$-wave RVB 
phase\cite{Iva00} and in the exact ground state of a small $t$-$J$ 
cluster.\cite{Leu00} Upon doping the SU(2) symmetry is
broken leading to two distinct phases, a $d$-wave RVB superconductor 
and a doped SF phase, a candidate for the pseudogap phase,\cite{DDW,flux} 
characterized by the opening of an antinodal gap in the excitation 
spectrum. Indeed, coexistence of sharp nodal quasiparticles and 
broad antinodal excitations have been found in angle-resolved photoemission 
spectroscopy (ARPES) studies on an array of underdoped cuprates such as 
La$_{2-x}$Sr$_x$CuO$_4$ (LSCO),\cite{Has07} 
Bi$_2$Sr$_2$CaCu$_2$O$_{8+\delta}$ (Bi2212),\cite{Tan06} and 
Ca$_{2-x}$Na$_x$CuO$_2$Cl$_2$ (Na-CCOC).\cite{Shen05}

However, there are also some low-$T$ properties of the SC state
which cannot be simply explained within the original RVB
framework. For example, following its theoretical
prediction,\cite{Zaa89} static charge and spin stripe
order has been detected in neutron scattering experiments and 
resonant soft $x$-ray scattering in some cuprate compounds as 
Nd-LSCO \cite{ndlsco} and La$_{2-x}$Ba$_x$CuO$_4$ (LBCO).\cite{lbco} 
More microscopic evidences of inhomogeneities have been given recently 
by scanning tunneling microscopy (STM) probing the doped-hole charge 
density with atomic resolution.\cite{Koh07} 
In the SC regime of two different cuprate families, Na-CCOC and Dy-Bi2212, 
\emph{bond-centered} charge patterns with a width of four lattice spacings 
have been seen as well as with two distinct types of spectral gaps:
(i) a relatively small one on arrays of two
neighboring chains (or ``ladder'') with clear coherence peaks
as expected for a $d$-wave superconductor, and (ii) a broader
pseudogap-like one on the other separating ladders.
This implies inhomogeneous superconductivity across the stripe unit cell.
Moreover, spin-glass behavior has been seen in the charge ordered state
by muon spin rotation\cite{Ohi05} with a characteristic
length of 2~nm, matching perfectly the STM data.\cite{Koh07} 
Also puzzling is the linear vanishing of the density of states at four 
nodal points that has been observed in the normal phase of LBCO at doping
$x=1/8$.\cite{Valla} 

From the above discussion, spatially modulated charge order coexisting
either with inhomogeneous AF or SC correlations appears to be ubiquitous 
among various families of underdoped cuprates. While inhomogeneous AF order
was predicted in the "stripe" scenario,\cite{Zaa89} SC phase coexisting 
with charge order is a new fascinating issue. 
Spatially oscillating $d$-wave superconductivity appearing on top of 
an AF stripe modulation has been proposed recently.\cite{Ogata}
We provide here a different theoretical proposal
based on a RMFT designed to describe {\it quantum disordered}
states\cite{Vojta02} (no incommensurate AF order is then assumed).
Although checkerboard states were first proposed,\cite{Checkerboard} 
we focus here on modulated SC solutions. Our conclusions are supported 
by VMC data.\cite{Manuela}

\begin{figure}[t!]
\vspace*{-1.5em}
\begin{center}
\unitlength=0.18cm
\begin{picture}(48,24)
\put(5,13){\includegraphics[width=0.35\textwidth]{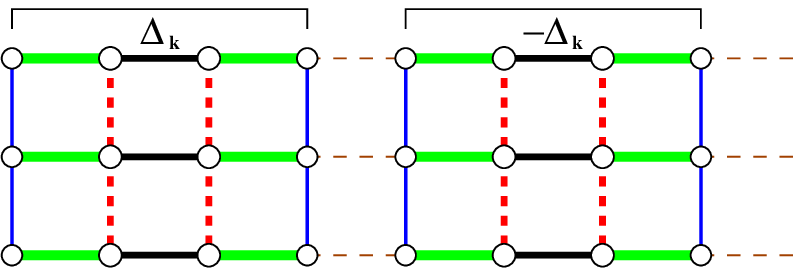}}
\put(5,0){\includegraphics[width=0.35\textwidth]{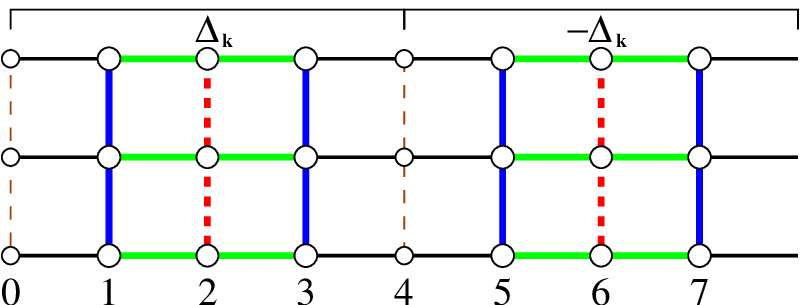}}
\put(0,21){ {\large (a)} } \put(0,8){ {\large (b)} }
\end{picture}
\end{center}
\caption {(color online) 
Spatial modulation of the hole density $n_{hi}$ (labeled from 0 to 7)
found at hole doping $x=1/8$ and $t/J=3$ for: (a) \emph{bond-}, and (b)
\emph{site-centered} $\pi$DRVB configurations. 
Circle diameters are proportional to the hole densities while widths
of bond lines connecting them are proportional to the magnitudes of the
pair-order parameter $\Delta_{ij}$.
}
\vspace*{-1.5em}
\label{fig:cart1}
\end{figure}

We start with a $t$-$J$ model:
\begin{equation}
{\cal H}= - t \sum_{\langle ij\rangle,\sigma}
     ({\tilde c}^{\dag}_{i\sigma}{\tilde c}^{}_{j\sigma} + h.c.)
      + J\sum_{\langle ij\rangle} {\bf S}_i \cdot {\bf S}_j,
\label{eq:H}
\end{equation}
where the sums run over the bonds.
Next, we replace the local constraints that restrict the fermion creation 
operators ${\tilde c}^{\dag}_{i\sigma}$ to the subspace with no doubly 
occupied sites by statistical Gutzwiller weights,\cite{Gut63} while 
decoupling in both particle-hole and particle-particle channels 
yields the following mean-field (MF) Hamiltonian,\cite{RVB,RMF}
\begin{align}
\label{eq:H_MF}
{\cal H}_{\rm MF}=&- t\sum_{\langle ij\rangle,\sigma} g_{ij}^t
                  (c^{\dagger}_{i,\sigma}c^{}_{j,\sigma}+h.c.)
                   -\mu\sum_{i,\sigma}n_{i,\sigma}\nonumber \\
                  &-\frac{3}{4} J \sum_{\langle ij\rangle,\sigma}g_{ij}^J
                  (\chi_{ji}c^{\dagger}_{i,\sigma}c^{}_{j,\sigma}
                  + h.c. -|\chi_{ij}|^2) \nonumber  \\
                  &-\frac{3}{4} J \sum_{\langle ij\rangle,\sigma}g_{ij}^J
                  (\Delta_{ji}c^{\dagger}_{i,\sigma}c^\dagger_{j,-\sigma}
                  + h.c. -|\Delta_{ij}|^2),
\end{align}
with the Bogoliubov-de Gennes self-consistency conditions for the bond-
$\chi_{ji}=\langle c^\dagger_{j,\sigma}c^{}_{i,\sigma}\rangle$ and pair-order
$\Delta_{ji}=\langle c^{}_{j,-\sigma}c^{}_{i,\sigma}\rangle
            =\langle c^{}_{i,-\sigma}c^{}_{j,\sigma}\rangle$
parameters in the unprojected state. Moreover, improved Gutzwiller weights
depending on local hole densities $n_{hi}=1-\sum_{\sigma}\langle
c^{\dagger}_{i,\sigma}c^{}_{i,\sigma}\rangle$ are used,
\begin{align}
\label{Eq:Gutz}
   g_{ij}^J &=\frac{4(1-n_{hi})(1-n_{hj})}
                   {\alpha_{ij}+8n_{hi}n_{hj}\beta_{ij}^{-}(2)
                   +16\beta_{ij}^{+}(4)}, \\
  g_{ij}^t &=\sqrt{\frac{4n_{hi}n_{hj}(1-n_{hi})(1-n_{hj})}
                   {\alpha_{ij}+8(1-n_{hi}n_{hj})|\chi_{ij}|^2
                   +16|\chi_{ij}|^4}},
\end{align}
with $\alpha_{ij}=(1-n_{hi}^2)(1-n_{hj}^2)$ and
$\beta_{ij}^{\pm}(n)= |\Delta_{ij}|^n\pm|\chi_{ij}|^n$, while the
$|\chi_{ij}|$ and $|\Delta_{ij}|$ terms account for the
correlations of the probabilities between nearest-neighbor
sites.\cite{RMF} Using unit cell translation symmetry,\cite{Rac06} 
calculations  (with hole doping $x=1/8$ and $t/J=3$) were carried out 
on a large $256\times 256$ cluster at a low temperature $\beta J=500$ 
eliminating finite size effects.

\begin{figure}[t!]
\vspace*{-1.5em}
\begin{center}
\unitlength=0.18cm
\begin{picture}(48,24)
\put(5,13){\includegraphics[width=0.35\textwidth]{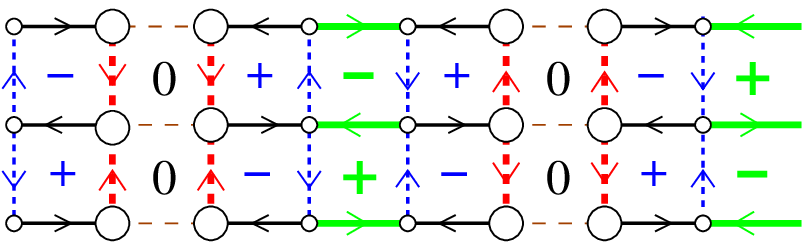}}
\put(5,0){\includegraphics[width=0.35\textwidth]{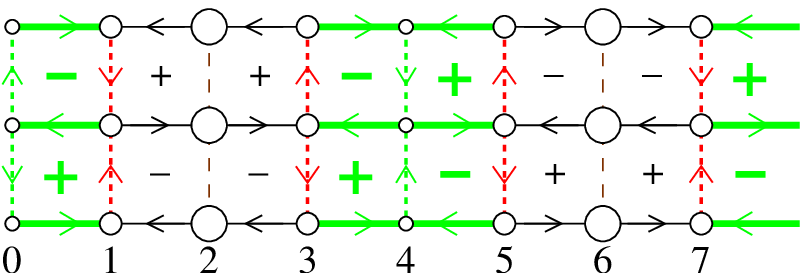}}
\put(0,21){ {\large (a)} } \put(0,8){ {\large (b)} }
\end{picture}
\end{center}
\caption {(color online) 
Same as in Fig.~\ref{fig:cart1} but for the $\pi$DSF patterns.
Bond lines with arrows indicate the direction of charge currents and
their widths are proportional to the magnitudes of the bond-order parameter
$\chi_{ij}$; $+/-$ symbols refer to positive/negative flux flowing
through each plaquette, the magnitude of which is represented by the size
of $+/-$ symbol.}
\vspace*{-1.5em}
\label{fig:cart2}
\end{figure}

The two modulated RVB states derived from the parent $d$-wave RVB phase 
found in this study are shown schematically in Fig.~\ref{fig:cart1}.
Hereafter we refer to them as $\pi$-phase domain RVB ($\pi$DRVB), as they
both involve two out-of-phase SC domains (see also Ref.~\onlinecite{Ogata}),
separated by horizontal/vertical bonds with vanishing pairing amplitudes, 
named as ``domain wall'' (DW), where $\Delta_{ij}$ gains a phase
shift of $\pi$. Specifically, we have found: (i)
\emph{bond-centered} $\pi$DRVB, with a maximum of the hole density spread 
over two-leg ladders depicted in Fig.~\ref{fig:cart1}(a), and (ii) 
\emph{site-centered} $\pi$DRVB, where the hole rich regions are composed 
of three-leg ladders as shown in Fig.~\ref{fig:cart1}(b).
Interestingly, out-of-phase adjacent SC domains has been advocated
\cite{Ogata,Ber07} as responsible for some puzzling decoupling of the 
CuO$_2$ planes in stripe-ordered LBCO.\cite{Li07}

\begin{figure}[t!]
\centerline{\includegraphics*[width=7.cm]{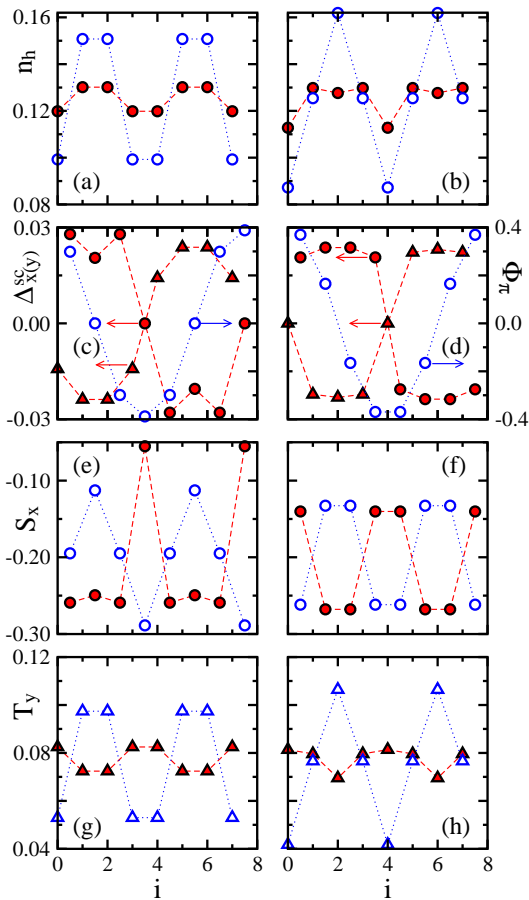}}
\caption {(color online)
(a,b) Hole density $n_{hi}$,
(c,d) SC order parameter $\Delta_{i\alpha}^{\rm SC}$ /
modulated flux $\Phi_{\pi i}$ indicated by left/right arrows, 
(e,f) spin correlation $S_{i}^{x}$, and
(g,h) bond charge $T_{i}^{y}$, found in the $\pi$DSF
(open symbols) and $\pi$DRVB (filled symbols) phase.
Left (right) panels depict \emph{bond-} (\emph{site-centered})
pattern; circles (triangles) in panels (c-h) correspond to the $x$ ($y$) 
direction, respectively.
}
\vspace*{-1.5em}
\label{fig:prof}
\end{figure}

Moreover, as shown in Fig.~\ref{fig:cart2}, we also found
closely related stripes originating from the SF state. An important 
characteristics of such patterns (named as $\pi$DSF) is again the 
existence of DWs which (i) act as nodes for the staggered current and 
(ii) introduce into the SF order parameter a phase shift of $\pi$. 
Hence, the $\pi$DSF phase could be regarded as quantum disordered 
AF stripes.\cite{Zaa89} In fact, it differs from the pair density wave
checkerboard solution proposed to interpret the pseudogap state, \cite{Chen04}
not only in terms of the spatial symmetry but also in that it breaks
time-reversal symmetry. 

Typical profiles of the site- and bond-centered normal $\pi$DSF 
and superconducting $\pi$DRVB phases are shown in 
Fig.~\ref{fig:prof} and the related MF energies are reported in 
Table~\ref{tab:1_8}, compared to VMC estimates.\cite{Manuela}
Here the locations of the DWs (oriented along the $y$-axis) of the SC and SF
states have been shifted by 2 lattice spacings (exactly as in
Figs.~\ref{fig:cart1} and \ref{fig:cart2}) for clarity.
Although MF results are mostly qualitative, as they are exact for particular
small clusters only,\cite{Oue07}
they are believed to correctly reflect the local spatial structure
of the various states, in particular the out-of-phase nature of the
respective order parameters 
$\Delta_{i\alpha}^{\rm SC}=g_{i,i+\alpha}^{t}\Delta_{i,i+\alpha}^{}$ 
and \emph{modulated} flux $\Phi_{\pi i}$\cite{Rac07} in neighboring domains as
shown in  Figs.~\ref{fig:prof}(c,d).
The hole density profiles in the bond-centered (site-centered)
stripes show the emergence of two (three) inequivalent
sites in the unit cell.
One important feature of these DWs is that they are
half-filled, i.e. the integrated charge density in the direction
perpendicular to the DWs corresponds
to an average of one hole every eight sites.
Although the large quantitative differences between the magnitudes
of the charge modulations (as well as their opposite signs)
in the superconducting and normal states might be
an artefact of the Gutzwiller/mean-field approximations,\cite{Manuela}
the generic trend towards charge segregation clearly reflects the 
competition between the superexchange energy ($E_J$) and the kinetic energy 
($E_t$) of doped holes.
For example, a reduction of the SC or flux order parameters 
(the latter known to frustrate coherent hole motion\cite{flux}) 
enables a large bond charge hopping 
$T_i^{y}=2g_{i,i+y}^{t}Re\{\chi_{i,i+y}\}$ along the DWs, 
as in the usual stripe scenario,\cite{Zaa89,Rac06}
at the expense of a reduction of the AF correlations
$S_i^{x}=-\frac{3}{2}g_{i,i+x}^{J}
(|\chi_{i,i+x}|^2+|\Delta_{i,i+x}|^2)$ along the transverse bonds.
Although earlier VMC studies~\cite{Iva04} suggested that the $d$-wave RVB 
state 
has a lesser tendency towards phase separation than the SF,
our results (including the VMC results~\cite{Manuela} of Table~\ref{tab:1_8})
clearly demonstrate that the modulated phases with commensurate period 
$\lambda=8$ (at $x=1/8$) are very competitive in energy with the homogeneous 
ones.\cite{Jastrow} Hence, under some circumstances, such as long-ranged 
Coulomb repulsion\cite{Checkerboard} or in-plane anisotropy, the $d$-wave 
stripe ordered phase might correspond to the global minimum as suggested 
by the experimental data for Nd-LSCO,\cite{ndlsco} LBCO,\cite{lbco} and
Na-CCOC,\cite{Koh07} all with a tetragonal structural distortion.
In addition, the reinforcement of the short-ranged AF correlations by the 
finite flux flowing through plaquettes of weakly doped regions in
the $\pi$DSF phase helps to optimize $E_J$. 
Consequently, such a phase is characterized by a better $E_J$ as compared to
the uniform SF state (see Table~\ref{tab:1_8}). 
Hence an increased tendency of the latter towards phase separation should 
result from increasing $J$ or reducing $x$, in which case the kinetic 
energy gain is relatively unimportant. 

\begin{figure}[t!]
\centerline{\includegraphics*[width=7.1cm]{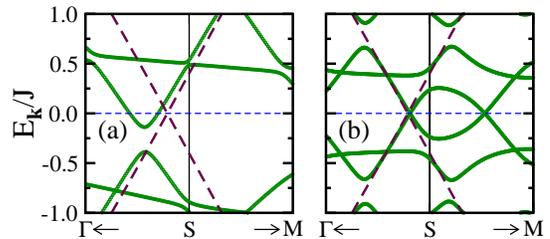}} \caption
{(color online) MF spectrum $E_{\bf k}$ of: (a) $\pi$DSF phase and
(b) $\pi$DRVB phase along the nodal \mbox{$\Gamma-M$} path 
\emph{near} the $S$ point. Dashed line shows $E_{\bf k}$ 
of the $d$-wave RVB phase. } 
\vspace*{-1.5em}
\label{fig:ek}
\end{figure}

\begin{table}[b!]
\caption {
MF kinetic energy $E_t$, magnetic energy $E_J$, and free energy $F$
as well as VMC energy $E_{\rm VMC}$ for a tilted cluster of 128 sites of 
the locally stable phases (all per site): $\pi$DSF, SF, 
\emph{bond-centered} $\pi$DRVB(1),
\emph{site-centered} $\pi$DRVB(2), and $d$-wave RVB one.
$\pi$DSF(1) and $\pi$DSF(2) phases are fully degenerate.
}
\begin{ruledtabular}
\begin{tabular}{cccccc}
 phase        & &    $E_t/J$   &    $E_J/J$    &    $F/J$  & $E_{\rm VMC}/J$ \cr
\colrule
 $\pi$DSF     & &    $-$0.8514 &   $-$0.4269   & $-$1.2783 & $-$1.3323 \cr
 SF           & &    $-$0.8622 &   $-$0.4230   & $-$1.2852 & $-$1.3389 \cr
 $\pi$DRVB(2) & &    $-$0.8736 &   $-$0.4491   & $-$1.3227 & $-$1.3359 \cr
 $\pi$DRVB(1) & &    $-$0.8719 &   $-$0.4518   & $-$1.3237 & $-$1.3359 \cr
 $d$-wave RVB & &    $-$0.8863 &   $-$0.4784   & $-$1.3647 & $-$1.3671 \cr
\end{tabular}
\end{ruledtabular}
\label{tab:1_8}
\end{table}

It is instructive to investigate the low-energy part of the MF
spectrum $E_{\bf k}$ (see Fig.~\ref{fig:ek}) which may reflect
interesting features of the true quasiparticle spectrum.
It is well known that the uniform SF phase has a cone-like
dispersion pinned at the $S=(\pi/2,\pi/2)$ point. However, doping
pushes the node above the chemical potential leading to small ``pockets'' 
and it has been recently shown that a diagonal $\pi$DSF phase follows 
the same trend.\cite{Rac07}
In contrast, since the vertical $\pi$DSF state mixes currents with 
different chiralities, one finds two cone-like features located symmetrically 
around the $S$ point. However, only one of them crosses the chemical 
potential which may lead to small electronic pockets weakly shifted towards
the Brillouin zone (BZ) center. They are reminiscent of the electron pockets 
obtained in a recent slave boson study of the $t-t'-U$ Hubbard
model.\cite{Rac07b} 
Interestingly, the parent $d$-wave RVB phase and its modulated $\pi$DRVB 
derivate both show a cone-like linear
quasiparticle excitation spectrum pinned at the chemical potential
at a location in the BZ close to the cone of the  vertical $\pi$DSF phase.
We also note that the presence of DWs induces the formation of a flat band 
in the vicinity of the antinodal direction that could be 
related to the ARPES observation in underdoped cuprates.
\cite{Has07,Tan06,Shen05}

\begin{figure}[t!]
\begin{center}
\setlength{\fboxsep}{0.1mm}
\unitlength=0.18cm
\begin{picture}(48,17)

\put(0, 0){\includegraphics*[width=8.6cm]{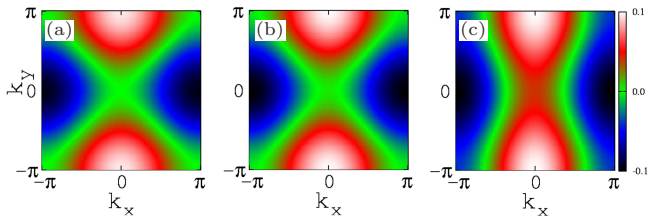}}

\put(3.15 ,13.6){\colorbox{white}{\scriptsize (a)}}
\put(18.65,13.6){\colorbox{white}{\scriptsize (b)}}
\put(33.75,13.6){\colorbox{white}{\scriptsize (c)}}

\end{picture}
\end{center}
\vspace*{-1.5em}
\caption {(Color) Real part of the SC order parameter 
$\Delta_{{\bf k}}^{\rm SC}({\bf Q})$ in: 
(a) $d$-wave RVB phase with ${\bf Q}=(0,0)$ as well as in the 
\emph{bond-centered} $\pi$DRVB phase  
with: 
(b) ${\bf Q}= \pm (\pi/4,0)$  and 
(c) ${\bf Q}= \pm (3\pi/4,0)$ (scaled resp. by factors 4.5 and 8 for clarity).  
}
\vspace*{-1.5em}
\label{fig:del}
\end{figure}

Lastly, in Figs.~\ref{fig:del}(a,b) we compare the (double) Fourier transform
of the SC order parameter $\Delta_{{\bf k}}^{\rm SC}({\bf Q})=
\frac{1}{N}\sum_{\langle ij\rangle}e^{i{\bf Q}\cdot{\bf R}_i
+i({\bf k}-{\bf Q}/2)\cdot({\bf R}_i-{\bf R}_j)}
\Delta_{ij}^{\rm SC}$
to the uniform RVB SC gap.
At the Cooper pair momenta ${\bf Q}=\pm (\pi/4,0)$
both real and imaginary parts (not shown) 
of $\Delta_{{\bf k}}^{\rm SC}({\bf Q})$ retain much of the $d$-wave symmetry
with a gap close to the nodal directions.
Moreover, in contrast to Ref.~\onlinecite{Ogata},
$\Delta_{{\bf k}}^{\rm SC}({\bf Q})$ is also non-vanishing in the
$\pi$DRVB phase for ${\bf Q}= \pm (3\pi/4,0)$.
Of lower amplitude the latter clearly reflects the DW structure through its deviation from $d$-wave symmetry, as seen in Fig.~\ref{fig:del}(c). 

To conclude, we note that the relative stability of the modulated
$d$-wave RVB superconductor should be highly sensitive to small 
perturbations of the microscopic Hamiltonian and, hence, can be relevant 
to some experimental cases such as the bond-centered electronic glass 
with unidirectional DWs in Na-CCOC and Dy-Bi2212 seen by STM.\cite{Koh07}
In fact, an experimental support to our modulated bond-centered 
superconducting state is provided by its strong similarities to the 
experimental pattern indicating the spontaneous appearance of 
two types of parallel ladders on nano-domains.\cite{Koh07}

\begin{acknowledgments}
We thank J. C. Davis and J. M. Tranquada for helpful correspondence.
M. Raczkowski acknowledges support from the Foundation for Polish Science
(FNP) and thanks the Laboratoires CRISMAT in Caen and IRSAMC in 
Toulouse for hospitality. 
This research was supported by the Polish Ministry of Science
and Education under Project No. N202 068 32/1481.
M. Capello thanks the Agence Nationale de la Recherche (France) for support.
\end{acknowledgments}


\begin{thebibliography}{00}

\bibitem{Voj00}  M. Vojta, Y. Zhang, and S. Sachdev,
                 \prb {\bf 62}, 6721 (2000).

\bibitem{tJ}     K. A. Chao, J. Spa\l{}ek, and A. M. Ole\'s,
                 J. Phys. C {\bf 10}, L271 (1977);
                 \prb \textbf{18}, 3453 (1978); 
                 F. C. Zhang and T. M. Rice,
                 {\it ibid.\/} {\bf 37},  3759 (1988).

\bibitem{Suc03}  S. Sachdev,
                 \rmp {\bf 75}, 913 (2003).

\bibitem{And87}  P. W. Anderson, Science {\bf 235}, 1196 (1987).

\bibitem{RVB}    F. C. Zhang {\it et al.},
                 Supercond. Sci. Technol. {\bf 1}, 36 (1988).

\bibitem{VMC}    C. Gros,
                 \prb {\bf 38}, 931 (1988); 
                 S. Sorella {\it et al.\/},
                 \prl {\bf 88}, 117002 (2002).

\bibitem{dwave}  G. Kotliar and J. Liu,
                 \prb {\bf 38}, 5142 (1988).

\bibitem{th_exp}
                 M. Calandra and S. Sorella, \prb {\bf 61}, R11894 (2000);
                 P. W. Anderson {\it et al.},
                 J. Phys. Condens. Matter \textbf{16}, R755 (2004).

\bibitem{sfp}    I. Affleck and J. B. Marston,
                 \prb {\bf 37}, R3774 (1988).

\bibitem{Iva00}  D. A. Ivanov, P. A. Lee, and X.-G. Wen,
                 \prl {\bf 84}, 3958 (2000).

\bibitem{Leu00}  P. W. Leung,
                 \prb {\bf 62}, R6112 (2000).

\bibitem{DDW}    S. Chakravarty {\it et al.},
                 \prb {\bf 63}, 094503 (2001).

\bibitem{flux}  For numerical computations of doped SF see, e.~g., 
                D. Poilblanc and Y. Hasegawa,
                \prb {\bf 41}, 6989 (1990);
                T. K. Lee and L. N. Chang,
                {\it ibid.\/} {\bf 42}, 8720 (1990).

\bibitem{Has07}  M. Hashimoto {\it et al.},
                 \prb {\bf 75}, 140503(R) (2007).

\bibitem{Tan06}  K. Tanaka {\it et al.},
                 Science {\bf 314}, 1910 (2006).

\bibitem{Shen05} K. M. Shen {\it et al.},
                 Science {\bf 307}, 901 (2005).

\bibitem{Zaa89} J. Zaanen and O. Gunnarsson,
                   \prb {\bf 40}, 7391 (1989);
                D. Poilblanc and T. M. Rice,
                   {\it ibid.\/} \textbf{39}, 9749 (1989);
                K. Machida,
                   Physica C \textbf{158}, 192 (1989).

\bibitem{ndlsco} J. M. Tranquada {\it et al.},
                 Nature (London) {\bf 375}, 561 (1995);
                 N. Christensen {\it et al.},
                 \prl {\bf 98}, 197003 (2007).

\bibitem{lbco}  M. Fujita {\it et al.},
                \prb {\bf 70}, 104517 (2004);
                P. Abbamonte {\it et al.},
                Nature Physics {\bf 1}, 155 (2005).

\bibitem{Koh07} Y. Kohsaka {\it et al.},
                Science {\bf 315}, 1380 (2007);
                see also J. Zaanen, {\it ibid.\/}
                {\bf 315}, 1372 (2007).

\bibitem{Ohi05} K. Ohishi {\it et al.},
                J. Phys. Soc. Jpn. {\bf 74}, 2408 (2005).

\bibitem{Valla} T. Valla {\it et al.},
                Science {\bf 314}, 1914 (2006).

\bibitem{Ogata} A. Himeda, T. Kato and M. Ogata, \prl {\bf 88}, 117001 (2002); 
                note that here a single sinusoidal modulation was assumed.

\bibitem{Vojta02} M. Vojta, \prb {\bf 66}, 104505 (2002).

\bibitem{Checkerboard} H.-X Huang, Y.-Q. Li, and F. C. Zhang,
                 \prb \textbf{71}, 184514 (2005);
                D. Poilblanc, {\it ibid.\/}  {\bf 72}, 060508(R) (2005);
                C. Weber {\it et al.}, {\it ibid.\/} {\bf 74}, 104506 (2006).

\bibitem{Manuela} The VMC results will be presented in detail elsewhere.

\bibitem{Gut63} D. Vollhardt,
                \rmp {\bf 56}, 99 (1984).

\bibitem{RMF}   M. Sigrist {\it et al.},
                \prb {\bf 49}, 12058 (1994).

\bibitem{Rac06} M. Raczkowski, R. Fr\'esard, and A. M. Ole\'s,
                \prb \textbf{73}, 174525 (2006);
                Europhys. Lett. \textbf{76}, 128 (2006).

\bibitem{Ber07} E. Berg {\it et al.}, \prl \textbf{99}, 127003 (2007).

\bibitem{Li07}  Q. Li {\it et al.}, \prl {\bf 99}, 067001 (2007).

\bibitem{Chen04} H.-D. Chen {\it et al.},  
                  \prl {\bf 93}, 187002 (2004).
\bibitem{Oue07} R. Fr\'esard, H. Ouerdane, and T. Kopp, (unpublished).

\bibitem{Rac07} M. Raczkowski {\it et al.},
                \prb {\bf 75}, 094505 (2007).
\bibitem{Iva04} D. A. Ivanov,
                \prb \textbf{70}, 104503 (2004).

\bibitem{Jastrow} Jastrow factors can lead to slightly lower energies
                  and to a change in the ranking of the various states, see
                  Ref.~\protect\onlinecite{Manuela}.
\bibitem{Rac07b} M. Raczkowski {\it et al.}, Phys. stat. sol. (b) {\bf 244},
  2521 (2007). 

\end{thebibliography}
\end{document}